\documentclass[twocolumn,floatfix,preprintnumbers,nofootinbib,superscriptaddress]{revtex4}

\usepackage{ulem}
\usepackage{bm}
\usepackage{times}
\usepackage{amssymb,amsbsy,amsmath,amsfonts}
\usepackage{graphicx}
\usepackage{float}
\usepackage{color}
\usepackage{morefloats}
\usepackage{rotating}
\usepackage{srcltx}
\usepackage{slashed}
\usepackage{subfigure}
\usepackage{multirow}
\usepackage{verbatim}
\usepackage{hyperref}
\usepackage{tabularx}
\usepackage{adjustbox}
\usepackage{array}
\usepackage{xcolor}

\begin{document}
	
	\title{Study $\Xi_c^+ \to \Lambda {\bar{K}}^0 \pi^+$  and search for the low-lying baryons $\Xi(1/2^-)$ and $\Sigma(1/2^-)$}
	 
	  \author{Ying Li}
   \affiliation{School of Physics, Zhengzhou University, Zhengzhou 450001, China}

\author{Wen-Tao Lyu}
\affiliation{School of Physics, Zhengzhou University, Zhengzhou 450001, China}

\author{Guan-Ying Wang}\email{wangguanying@henu.edu.cn}
\affiliation{School of Physics and Electronics, Henan University, Kaifeng 475004, China}

  \author{Longke Li}\email{lilongke@hunnu.edu.cn
}
   \affiliation{School of Physics and Electronics, Hunan Normal University, Changsha 410081, China}

   \author{Wen-Cheng Yan}\email{yanwc@zzu.edu.cn}
   \affiliation{School of Physics, Zhengzhou University, Zhengzhou 450001, China}
   
  
  \author{En Wang}\email{wangen@zzu.edu.cn}
\affiliation{School of Physics, Zhengzhou University, Zhengzhou 450001, China}

	\begin{abstract}
	Since searches for the low-lying excited baryons $\Xi(1/2^-)$ and $\Sigma(1/2^-)$ are crucial to deepening our understanding of the light baryon spectrum, we have investigated the Cabibbo-favored process $\Xi_c^+ \to \Lambda {\bar{K}}^0 \pi^+$ by taking into account the $S$-wave pseudoscalar meson-octet baryon interactions within the chiral unitary  approach, which could dynamically generate the resonances $\Xi(1/2^-)$ and $\Sigma(1/2^-)$. The contributions from the excited kaons are double Cabibbo-suppressed, and the contribution from the $\Sigma(1385)$ is also suppressed due to Korner-Pati-Woo theory, thus those states are expected to play negligible contributions in this process. We have predicted the $\bar{K}^0 \Lambda$ and $\pi^+\Lambda$ invariant mass distributions, which have the clear signals of the $\Xi(1/2^-)$ and $\Sigma(1/2^-)$. Thus, the $\Xi_c^+ \to \Lambda {\bar{K}}^0 \pi^+$ is an ideal process to search for the low-lying baryons $\Xi(1/2^-)$ and $\Sigma(1/2^-)$, and we make a call for a precise measurements of this process in experiments, such as Belle II, LHCb, and the proposed Super Tau-Charm Facility (STCF).	

	\end{abstract}
	
	\maketitle
	
	\section{Introduction} \label{sec:Introduction}

   	As we know, the light ground octet baryons and decuplet baryons have been well established, however there are some puzzles for the low-lying excited baryons with spin parity of $J^P=1/2^-$. One of the puzzles is the ``mass reverse problem'', i.e. the mass of the $N(1535)$ with spin-parity quantum numbers $J^P=1/2^-$ should be lower than the masses of the radial excitation $N(1440)$ with $J^P=1/2^+$ and the $\Lambda(1405)$ with strangeness $S=-1$ and $J^P=1/2^-$ within the naive quark model~\cite{Isgur:1978xj,Zou:2010tc,Zhang:2004xt}. Many theoretical works suggest that they should have more exotic structure than the three valence quarks composition, such as compact pentaquark, molecular state, or the mixing of different components\cite{Zou:2010tc,Zhang:2004xt,Zou:2007mk,Liu:2005pm}.

    	For the low-lying baryons with spin parity $J^P=1/2^-$, in addition to the $N(1535)$ and $\Lambda(1405)$, the other low-lying baryons $\Sigma(1/2^-)$ and $\Xi(1/2^-)$ have not been well established by now~\cite{Wang:2024jyk}. According to the 2024 edition of  Review of Particle Physics (RPP)~\cite{ParticleDataGroup:2024cfk}, there is a 1-star state $\Sigma(1620)$ with $J^P=1/2^-$, which implies that the existence of the $\Sigma(1620)$ is still in debate and needs to be confirmed with more precise measurements~\cite{Wang:2024jyk,Sarantsev:2019xxm}. Recently, based on the suggestion of Refs.~\cite{Xie:2017xwx,Lyu:2024qgc}, the BESIII Collaboration reported the evidence of the $\Sigma(1/2^-)$ with mass of 1380~MeV and the statistical significance is greater than $3\sigma$ in the process $\Lambda_c^+\to \Lambda \pi^+\eta$~\cite{BESIII:2024mbf}. Later, Ref.~\cite{Duan:2024led} has well reproduced the BESIII measured invariant mass distributions of this process by considering the contributions from the $a_0(980)$ and $\Lambda(1670)$, as the dynamically generated resonances from the meson-meson and meson-baryon interactions, and the contribution from the intermediate $\Sigma(1385)$.

On the other hand, the Belle Collaboration has observed significant cusp structure in the $\Lambda\pi^+$ and $\Lambda\pi^-$ invariant mass distributions of the process $\Lambda_c^+\to \Lambda\pi^+\pi^+\pi^-$~\cite{Belle:2022ywa}, which could be interpreted as a resonance with mass of $(1434.3\pm0.6\pm 0.9)$~MeV and width of $(11.5\pm2.8\pm 5.3)$~MeV for the $\Lambda\pi^+$ combination, and with mass of $(1438.5\pm0.9\pm 2.5)$~MeV and width of $(33.0\pm 7.5\pm 23.6)$~MeV for the $\Lambda\pi^-$ combination. Its average mass is consistent with the theoretically predicted mass of $\Sigma(1/2^-)$, which is dynamically generated from the $S$-wave meson-baryon interaction within the chiral unitary approach~\cite{Oset:1997it,Oller:2000fj,Roca:2013cca,Guo:2012vv,Jido:2003cb}.

For the $\Xi(1/2^-)$, there are two candidates listed in the RPP~\cite{ParticleDataGroup:2024cfk}, one is the 2-star state $\Xi(1620)$ with mass about $1620$~MeV and width of $(32^{+8}_{-9})$~MeV, and the other one is 3-star state $\Xi(1690)$ with mass of $(1690\pm 10)$~MeV and mass of $(20\pm 15)$~MeV. However, the spin-parity quantum numbers of these two states have not been determined~\cite{ParticleDataGroup:2024cfk}. In 2002, the Belle and FOCUS Collaborations have reported the evidence of the resonant contribution $\Lambda_c^+\to \Xi(1690)^0 K^+$~\cite{Belle:2001hyr,FOCUS:2005sye}. In 2005, the BaBar Collaboration has observed the $\Xi(1690)^0$ resonance in the decay $\Lambda_c^+\to \Lambda \bar{K}^0 K^+$ and measured the mass and width of this state to be $(1684.7\pm1.3(\mathrm{stat})^{+2.2}_{-1.6}(\mathrm{syst}))$~MeV and $(8.1^{+3.9}_{-3.5}(\mathrm{stat})^{+1.0}_{-0.9}(\mathrm{syst}))$~MeV, with favored spin $J=1/2$~\cite{BaBar:2006tck}. Subsequently, in the process $\Lambda_c^+\to \Xi^-\pi^+ K^+$, the BaBar Collaboration has also indicated that the spin-parity quantum numbers are $J^P=1/2^-$ for the $\Xi(1690)$~\cite{BaBar:2008myc}. In 2019, the Belle Collaboration has reported the first observation of the double strange baryon $\Xi(1620)^0$ through the process $\Xi_c^+\to\Xi^-\pi^+\pi^+$, with the significance of $4.0\sigma$~\cite{Belle:2018lws}. Subsequently, the Belle Collaboration has analyzed this process and found the $\Xi(1620)^0$ and $\Xi(1690)^0$ in the $\Xi\pi$ decay mode~\cite{Sumihama:2020mqa}. There are some predictions for the $\Xi(1/2^-)$ within the quark models~\cite{Capstick:1986ter,Glozman:1995fu,Melde:2008yr}. For instance, within the quark models the predicted mass of $\Xi(1/2^-)$ is approximately 1800~MeV~\cite{Chao:1980em,Capstick:1986ter}, larger than the masses of the $\Xi(1620)^0$ and $\Xi(1690)^0$. 	With a large $N_c$ QCD approach, Ref.~\cite{Schat:2001xr} has obtained the $\Xi(1/2^-)$ state with mass of 1779~MeV, while the Skyrme model predicted two $\Xi(1/2^-)$ states with masses of 1616~MeV and 1658~MeV, respectively~\cite{Oh:2007cr}. 
In Refs.~\cite{Pervin:2007wa,Xiao:2013xi},  the $\Xi(1690)$ was regarded as the first orbital excitation with $J^P=1/2^-$.  In Refs.~\cite{Ramos:2002xh,Garcia-Recio:2003ejq}, both the $\Xi(1620)$ and $\Xi(1690)$ appeared with $J^P=1/2^-$, and were dynamically generated from
the interaction of the coupled channels $\pi\Xi$, $\bar{K}\Lambda$, $\bar{K}\Sigma$, and $\eta\Xi$.

Therefore, the experimental determination of the $\Sigma(1/2^-)$ and $\Xi(1/2^-)$ remains necessary, and the searches for these low-lying baryons, $\Sigma(1/2^-)$ and $\Xi(1/2^-)$, are crucial for our understanding of the properties of low-lying baryons with $J^P = 1/2^-$. 
In Refs.~\cite{Wang:2015qta,Liu:2017hdx}, it was suggested to search for the $\Sigma(1/2^-)$ in the processes $\chi_{c0}\to \bar\Sigma \Sigma \pi$ and $\chi_{c0}\to \bar\Lambda \Sigma \pi$. The photoproduction of the $\Sigma(1/2^-)$ is also proposed in Refs.~\cite{Kim:2021wov,Lyu:2023oqn}. Recently, Ref.~\cite{Li:2024rqb} has investigate the process $\Lambda_c^+\to p\bar{K}^0\eta$, and found the invariant mass distributions measured by Belle~\cite{Belle:2022pwd} could be well described by considering the contribution from the predicted $\Sigma(1/2^-)$.  In addition, the state $\Sigma(1/2^-)$ is also expected to be observed in the processes $\Lambda_c^+\to \pi^+\pi^0\pi^-\Sigma^+$~\cite{Xie:2018gbi}, $\Lambda_c^+\to \gamma\pi^+\Lambda$~\cite{Wang:2024ewe}.
In Ref.~\cite{Li:2024tvo}, the
correlation functions for the meson-baryon scattering are studied in the chiral unitary approach, and an excited $\Sigma(1/2^-)$ state is predicted around the $\bar{K}N$ threshold.
Meanwhile, it is suggested to search for the $\Xi(1690)^-$ in the reactions of $K^-p\to K^+K^-\Lambda$ by Refs.~\cite{Ahn:2018hbc,Nam:2019mon}, $\Xi _c^+ \rightarrow \Xi ^- \pi ^+ \pi ^+$~\cite{Li:2023olv}, and $\Lambda^0_b \to J/\psi p \pi^-$~\cite{Wang:2015pcn}.

It should be pointed out that,	the low-lying baryons $\Lambda(1405)$ and $N(1535)$ could be interpreted as dynamically generated state of the meson-baryon interaction within the chiral unitary approach. Within the same framework, $\Sigma(1/2^-)$ and $\Xi(1/2^-)$ are also expected to be dynamically generated from the meson-baryon interaction, since the predicted $\Sigma(1/2^-)$ with mass around $\bar{K}N$ threshold is confirmed by the Belle's measurements of the process $\Lambda^+_c\to \Lambda\pi^+\pi^+\pi^-$. Meanwhile,  Ref.~\cite{Ramos:2002xh} has studied the low-energy meson-baryon scattering in the strangeness $S=-2$ sector within the chiral unitary approach and found a scattering-matrix pole around 1605~MeV which corresponds to the $\Xi(1620)$ quoted by the RPP, and Refs.~\cite{Wang:2019krq,Chen:2019uvv,Huang:2020taj} also investigated the molecular picture of the $\Xi(1620)$ as a $\Lambda\bar{K}$ or $\Sigma\bar{K}$ bound state. In Refs.~\cite{Garcia-Recio:2003ejq,Gamermann:2011mq}, both the $\Xi(1620)$ and  $\Xi(1690)$ were dynamically generated with $J^P=1/2^-$, resulting from the interaction of the coupled channels $\pi\Xi$, $\bar{K}\Lambda$, $\bar{K}\Sigma$, and $\eta\Xi$, and a notable observation was that the $\Xi(1690)$ exhibited strong coupling with $\bar{K}\Sigma$ and weak coupling with $\pi\Xi$, while the $\Xi(1620)$ exhibited the opposite behavior. Recently, by fitting the Belle's measurements on the processes $\Lambda_c^+\to\Lambda K^+\bar{K}^0$ and $\Xi_c^+\to\Xi^-\pi^+\pi^+$, Refs.~\cite{Liu:2023jwo,Magas:2024mba} have examined the molecular nature of the $\Xi(1690)$ state, and concluded that those experimental data support the molecular state interpretation of the $\Xi(1690)$.
Thus, search for the predicted  $\Sigma(1/2^-)$ and $\Xi(1/2^-)$ is crucial to shed light on the properties of the low-lying baryons.

Recently, the BESIII, Belle/Belle II, and LHCb Collaborations have accumulated lots of information on the hadronic decays of the charmed baryons~\cite{BESIII:2024xny,BESIII:2022udq,Belle:2022ywa,Belle:2022pwd,Ryzka:2023wjw}, which have been used to investigate the light hadrons~\cite{Zhang:2024jby,Wang:2022nac,Feng:2020jvp,Wang:2020pem,Li:2025gvo,Zeng:2020och}. Thus, we propose to simultaneously search for the $\Sigma(1/2^-)$ and $\Xi(1/2^-)$ states in the Cabibbo-favored process $\Xi_c^+ \to \Lambda {\bar{K}}^0 \pi^+$. In this process, the contribution from the excited kaons is double Cabibbo-suppressed, and thereby is negligible. Meanwhile, the contribution from the $\Sigma(1385)$ is also small due to the Korner-Pati-Woo theory and $\Xi_c-\Xi'_c$ mixing. Indeed, in Ref.~\cite{Miyahara:2016yyh}, the authors have also pointed out that this process $\Xi^+_c\to \Lambda\bar{K}^0\pi^+$ is less influenced by the possible meson resonances.

    In this work, we will investigate the process  $\Xi_c^+ \to \Lambda {\bar{K}}^0 \pi^+$ by taking into account the contributions from the meson-baryon interaction within the chiral unitary approach~\cite{Roca:2013cca,Sekihara:2015qqa}, which could dynamically generate the $\Sigma(1/2^-)$ and $\Xi(1/2^-)$, and make the predictions for their possible signals in the invariant mass distributions and the Dalitz plot.

    The structure of this paper is as follows. In Sec.~\ref{sec:Formalism}, we present the
    theoretical formalism of the process $\Xi_c^+\to\Lambda\bar{K}^0\pi^+$, and
    in Sec.~\ref{sec:Results}, we show our numerical results and discussions, 
    followed by a short summary in the last section.

	\section{Formalism} \label{sec:Formalism}

In this section, we will introduce the theoretical formalism for the process $\Xi_c^+ \to \Lambda {\bar{K}}^0 \pi^+$. First, the mechanism for this process via the intermediate state $\Xi(1/2^-)$ is presented in Sec.~\ref{susec:A}, and the mechanism via the intermediate state $\Sigma(1/2^-)$ is presented in Sec.~\ref{susec:B}. Finally, we give the formalism for the invariant mass distributions of this process in Sec.~\ref{susec:C}.

	\subsection{Mechanism of the intermediate $\Xi(1/2^-)$}\label{susec:A}

		\begin{figure}[tbhp]
		\centering
		\includegraphics[scale=0.6]{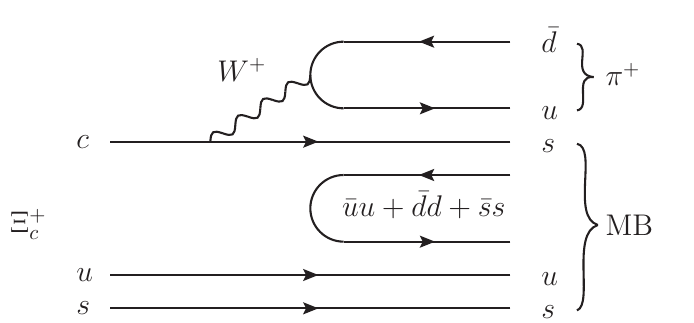}
		\caption{Quark-level diagram for the Cabibbo-favored process $\Xi_c^+\to\pi^+MB$.}
		\label{fig:quark1690}
	\end{figure}
	
\begin{figure}[tbhp]
		\centering
		\includegraphics[scale=0.6]{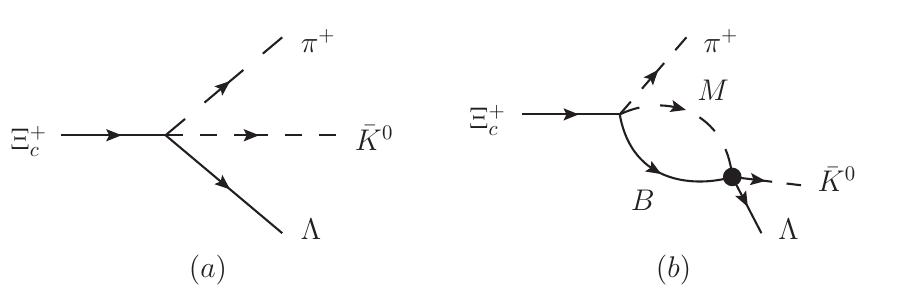}
		\caption{Tree-level diagram (a) and the $S=-2$ final state interaction diagram (b) for the process $\Xi_c^+\to\Lambda\bar{K}^0\pi^+$.}
		\label{fig:hadron1690}
	\end{figure}

In this work, we consider the diagram of the $W^+$ external emission for the process $\Xi_c^+ \to \Lambda {\bar{K}}^0 \pi^+$, as shown in Fig.~\ref{fig:quark1690}. In the first step, the charmed quark of the initial $\Xi_c^+$ weakly decays into a strange quark and a $W^+$ boson, followed by the $W^+$ boson decaying into the $u\bar{d}$ pair. In the second step, the $u\bar{d}$ pair from the $W^+$ boson hadronizes into $\pi^+$, and the $s$ quark and the $us$ quarks of the initial $\Xi^+_c$, together with a $\bar{q}q$ pair with vacuum quantum numbers created from the vacuum, hadronize into a pair of meson and baryon, as expressed as
	\begin{align}
		\Xi_c^+&\Rightarrow \frac{1}{\sqrt{2}}c\left(su-us\right)\nonumber\\
        &\Rightarrow\frac{1}{\sqrt{2}}u\bar{d}s\left(\bar{u}u+\bar{d}d+\bar{s}s\right)\left(su-us\right)\nonumber \\
        &\Rightarrow\frac{1}{\sqrt{2}}\pi^+ s\left(\bar{u}u+\bar{d}d+\bar{s}s\right)\left(su-us\right),
	\end{align}
	where we take the flavor wave function $\Xi_c^+=\frac{1}{\sqrt{2}}c\left(su-us\right)$. 
Here we use the $\mathrm{SU}(3)$ pseudoscalar mesons matrix $M$ and the baryons matrix $B$ to connect two degrees of freedom, the quarks and the hadrons, based on the SU(3) flavor symmetry. The explicit forms of the matrix $M$ and $B$ are given as follows~\cite{Miyahara:2016yyh,Lyu:2023aqn}
		\begin{eqnarray}
		M&=&\left(\begin{matrix} u\bar{u} & u\bar{d} & u\bar{s}  \\
			d\bar{u}  &   d\bar{d}  &  d\bar{s} \\
			s\bar{u}  &  s\bar{d}   &    s\bar{s}
		\end{matrix}
		\right)      \nonumber  \\
        \qquad\nonumber  \\
	&=&\left(\begin{matrix} \frac{\eta}{\sqrt{3}}+ \frac{{\pi}^0}{\sqrt{2}}+ \frac{{\eta}'}{\sqrt{6}} & \pi^+ & K^+  \\
			\pi^-  &   \frac{\eta}{\sqrt{3}}- \frac{{\pi}^0}{\sqrt{2}}+ \frac{{\eta}'}{\sqrt{6}}  &  K^0 \\
			K^-  &  \bar{K}^{0}   &    -\frac{\eta}{\sqrt{3}}+ \frac{{\sqrt{6}\eta}'}{3}
		\end{matrix}
		\right),\nonumber\\
		B &=& \frac{1}{\sqrt{2}}\begin{pmatrix} 
			u(ds-sd) & u(su-us) & u(ud-du)  \\
			d(ds-sd) & d(su-us) & d(ud-du) \\
			s(ds-sd) & s(su-us) & s(ud-du)
		\end{pmatrix} \nonumber \\
        \qquad\nonumber  \\
		&=& \begin{pmatrix} 
			\frac{\Sigma^0}{\sqrt{2}} + \frac{{\Lambda}}{\sqrt{6}} & \Sigma^+ & p  \\
			\Sigma^- & -\frac{\Sigma^0}{\sqrt{2}} + \frac{{\Lambda}}{\sqrt{6}}  & n \\
			\Xi^-  & \Xi^{0} & -\frac{2\Lambda}{\sqrt{6}} 		\end{pmatrix},
		\label{eq:MB}
	\end{eqnarray}
	where we have considered the $\eta$-$\eta^{\prime}$ mixing~\cite{Bramon:1992kr,Lyu:2023ppb}. With the mesons matrix $M$ and baryons matrix $B$, we could express the hadronization as
     \begin{align}
     		\Xi_c^+      \Rightarrow&\frac{1}{\sqrt{2}}\pi^+ s\left(\bar{u}u+\bar{d}d+\bar{s}s\right)\left(su-us\right) \nonumber \\
=&\pi^+\sum M_{3i}B_{i2}\nonumber\\
		=&\pi^+\left(|K^-\Sigma^+\rangle-\frac{1}{\sqrt{2}}|\bar{K}^0\Sigma^0\rangle \right. \nonumber \\ &\left.+\frac{1}{\sqrt{6}}|\bar{K}^0\Lambda\rangle-\frac{1}{\sqrt{3}}|\eta\Xi^0\rangle \right),
		\label{eq:1690MBchannel}
	\end{align}
where the final state $\Lambda\bar{K}^0\pi^+$ could be produced directly, as shown in Fig.~\ref{fig:hadron1690}(a). Meanwhile, the final state $\Lambda\bar{K}^0\pi^+$ could be achieved through the transition of the $\bar{K}\Sigma/\eta\Xi/\bar{K}\Lambda\to \bar{K}\Lambda$, as shown in Fig.~\ref{fig:hadron1690}(b). Thus, we can write down the amplitude of the process $\Xi_c^+\to\Lambda\bar{K}^0\pi^+$ for the contribution of Fig.~\ref{fig:hadron1690} as
	\begin{align}
	\mathcal{T}^{\Xi(1/2^-)}=&CV_p\left[\frac{1}{\sqrt{6}}+G_{K^-\Sigma^+}(M_{\bar{K}^0\Lambda})t_{K^-\Sigma^+ \to \bar{K}^0\Lambda}(M_{\bar{K}^0\Lambda})\right. \nonumber \\
	&\left.-\frac{1}{\sqrt{2}}G_{\bar{K}^0\Sigma^0}(M_{\bar{K}^0\Lambda})t_{\bar{K}^0\Sigma^0 \to \bar{K}^0\Lambda}(M_{\bar{K}^0\Lambda})\right.
	\nonumber \\
	&\left.+\frac{1}{\sqrt{6}}G_{\bar{K}^0\Lambda}(M_{\bar{K}^0\Lambda})t_{\bar{K}^0\Lambda \to \bar{K}^0\Lambda}(M_{\bar{K}^0\Lambda})\right.
	\nonumber \\
	&\left.-\frac{1}{\sqrt{3}}G_{\eta\Xi^0}(M_{\bar{K}^0\Lambda})t_{\eta\Xi^0 \to \bar{K}^0\Lambda}(M_{\bar{K}^0\Lambda})\right],   	\label{eq:T1690}
\end{align}
     where the color factor $C$ represents the relative weight of the external $W^+$ excitation mechanism shown in Fig.~\ref{fig:quark1690} with respect to the internal $W^+$ excitation mechanism described later in Fig.~\ref{fig:quark1430}, and the parameter $V_p$ contains the dynamical information of the weak interaction vertex shown in Fig.~\ref{fig:quark1690},and is assumed to be constant in the relevant energy region.\footnote{Indeed, the factor $V_p$ contains the transition form factors, and should smoothly depend on the invariant mass of $\bar{K}^0\Lambda$ and $\pi\Lambda$~\cite{Geng:2024sgq}. In this work, we mainly focus on the final state interactions, which produce the $\Xi(1/2^-)$ and $\Sigma(1/2^-)$ resonances, and one expects that the assumption of a constant $V_p$ should not modify the peak structure of these two states significantly.} The coefficients of each term represent the weight of the coupling channels, are obtained from Eq.~\eqref{eq:1690MBchannel}. $G$ is the meson-baryon loop function, which can be regularized using either the three-momentum cutoff method or dimensional regularization. In this work, we choose the latter method, and the corresponding analytic form of the loop function for the $MB$ channel, $G_{MB}$ is given by 
 	\begin{align}
		G_{MB} &= \frac{2M_B}{16\pi^2} \Bigg\{ a_{MB}(\mu) + \ln\frac{M_B^2}{\mu^2} + \frac{m_M^2 - M_B^2 + s}{2s} \ln\frac{m_M^2}{M_B^2} \nonumber \\
		&\quad + \frac{q_{MB}}{\sqrt{s}} \bigg[ \ln\left(s - M_B^2 + m_M^2 + 2q_{MB}\sqrt{s}\right) \nonumber \\
		&\quad + \ln\left(s + M_B^2 - m_M^2 + 2q_{MB}\sqrt{s}\right) \nonumber \\
		&\quad - \ln\left(-s + M_B^2 - m_M^2 + 2q_{MB}\sqrt{s}\right) \nonumber \\
		&\quad - \ln\left(-s - M_B^2 + m_M^2 + 2q_{MB}\sqrt{s}\right) \bigg] \Bigg\},
			\label{eq:G}
	\end{align}
	where $a_{MB}(\mu)$ is the subtraction constant, and $\mu$ is a scale of dimensional regularization.  $m_M$ and $M_B$ represent the masses of the meson and baryon, respectively. $q_{MB}$ is the on-shell meson momentum in the center-of-mass frame of the meson-baryon system, given by $q_{MB}=\lambda^{1/2}(s,m_M^2,M_B^2)/2\sqrt{s}$ 
with $\lambda(x,y,z)=x^2+y^2+z^2-2xy-2yz-2zx$.
	
		\begin{table*}[htbp]
		\centering
		\caption {$C_{ij}$ coefficients in the potential for the coupled channels with $S=-2$ ($C_{ij}$=$C_{ji}$)~\cite{Sekihara:2015qqa}.}	\label{tab:I}
		\begin{tabular}{lcccccc}
			\hline\hline  
			\qquad\quad&$K^-\Sigma^+$\qquad\quad&$\bar{K}^0\Sigma^0$\qquad\quad&$\bar{K}^0\Lambda$\qquad\quad&$\pi^+\Xi^-$\qquad\quad&$\pi^0\Xi^0$\qquad\quad&$\eta\Xi^0$\\  \hline 
			$K^-\Sigma^+$\qquad\qquad&1\qquad\qquad &$-\sqrt{2}$ \qquad\qquad &0\qquad\quad&0 &$-1/\sqrt{2}$ \qquad\qquad &$-\sqrt{3/2}$\\ $\bar{K}^0\Sigma^0$\qquad\qquad&\qquad\qquad\qquad &0\qquad\qquad&0\qquad\quad&$-1/\sqrt{2}$\qquad\qquad &-1/2\qquad\qquad &$\sqrt{3/4}$\\ $\bar{K}^0\Lambda$\qquad\qquad&\qquad\qquad\qquad&\qquad\qquad\qquad &0\qquad\quad&$-\sqrt{3/2}$\qquad\qquad &$\sqrt{3/4}$\qquad\qquad &-3/2   \\
			$\pi^+\Xi^-$\qquad\qquad&\qquad\qquad\qquad&\qquad\qquad\qquad &\qquad\qquad\quad&1\qquad\qquad &$-\sqrt{2}$\qquad\qquad &0   \\		
			$\pi^0\Xi^0$\qquad\qquad&\qquad\qquad\qquad&\qquad\qquad\qquad &\qquad\qquad\quad&\qquad\qquad\qquad &0\qquad\qquad &0   \\		
			$\eta\Xi^0$\qquad\qquad&\qquad\qquad\qquad&\qquad\qquad\qquad &\qquad\qquad\quad&\qquad\qquad\qquad &\qquad\qquad\qquad &0   \\	
			\hline\hline
		\end{tabular}
	\end{table*}

    The $t_{MB\to \bar{K}^0\Lambda}$ in Eq.~\eqref{eq:T1690} represents the transition amplitudes between $MB$ channel and $\bar{K}^0\Lambda$ channel, which can be obtained by solving the Bethe-Salpeter equation
    \begin{eqnarray}
    	T=[1-VG]^{-1}V,
    \end{eqnarray}
    with the meson-baryon loop function $G$ shown in Eq.~\eqref{eq:G} and the interaction kernel $V_{ij}$ taken from the chiral Lagrangians, which is expressed as
		\begin{align}
		V_{ij}=&-C_{ij}\frac{1}{4f_if_j}(2\sqrt{s}-M_i-M_j)\nonumber\\
		&\sqrt{\frac{E_i+M_i}{2M_i}}\sqrt{\frac{E_j+M_j}{2M_j}},
	\end{align} 
	where $M_{i(j)}$ are the masses of the baryons in the $i(j)$-th channel, and the $E_{i(j)}$  are the energies of the corresponding baryons with the expression $E_i=(s+M^2_i-m^2_i)/2\sqrt{s}$. Here we take into account the coupled channels, $K^-\Sigma^+$, $\bar{K}^0\Sigma^0$, $\bar{K}^0\Lambda$, $\pi^+\Xi^-$, $\pi^0\Xi^0$, and $\eta\Xi^0$, and the coefficients $C_{ij}$  shown in Table~\ref{tab:I} are taken from Ref.~\cite{Sekihara:2015qqa} and reflect the 
SU(3) flavor symmetry. The coefficients $f_i$ are the pseudoscalar decay constants, and we use,
	\begin{eqnarray}
		f_\pi=92.4~\mathrm{MeV},\qquad f_K=1.123f_\pi,\qquad f_\eta=1.3f_\pi.
	\end{eqnarray}
In addition, we take $a_{\bar{K}\Sigma} =-1.98, a_{\bar{K}\Lambda} =-2.07, a_{\pi\Xi} =-0.75, a_{\eta\Xi} =-3.31$, and $\mu = 630$ MeV for the loop function $G$, as used in Ref.~\cite{Sekihara:2015qqa}.

	\subsection{Mechanism for the intermediate $\Sigma(1/2^-)$}\label{susec:B}
	
		\begin{figure}[tbhp]
		\centering
		\includegraphics[scale=0.6]{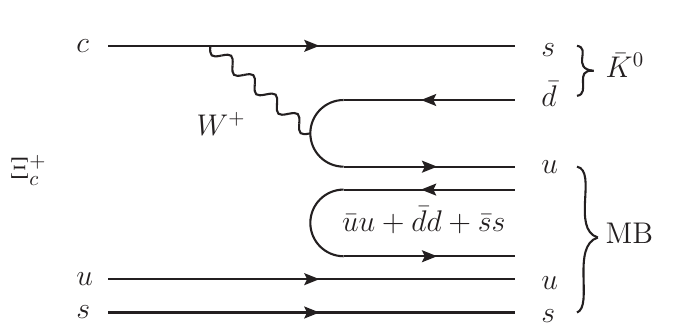}
		\caption{Quark-level diagram for the $\Xi_c^+\to\bar{K}^0MB$ process.}
		\label{fig:quark1430}
	\end{figure}
	
	In this subsection, we discuss the theoretical mechanism of the dynamic generation of $\Sigma(1/2^-)$ in the process $\Xi_c^+ \to \Lambda {\bar{K}}^0 \pi^+$, which can happen via the internal $W^+$ emission diagram, as shown in Fig.~\ref{fig:quark1430}. Firstly, the $c$ quark in the $\Xi_c^+$ decays into an $s$ quark and a $W^+$ boson via the weak interaction. Subsequently, the $W^+$ boson  decays weakly into a $u\bar{d}$ pair. Then, the $s\bar{d}$ quarks hadronize into $\bar{K}^0$, while the remaining quarks $uus$ cluster, together with the $\bar{q}q$ pair generated from the vacuum, hadronize into meson-baryon pairs, 
	which could be expressed as
	\begin{align}
		\Xi_c^+&\Rightarrow\frac{1}{\sqrt{2}}s\bar{d}u\left(\bar{u}u+\bar{d}d+\bar{s}s\right)\left(su-us\right)\nonumber\\
		&\Rightarrow \bar{K}^0\frac{1}{\sqrt{2}}u\left(\bar{u}u+\bar{d}d+\bar{s}s\right)\left(su-us\right). 
	\end{align}
\begin{figure}[tbhp]
		\centering
		\includegraphics[scale=0.6]{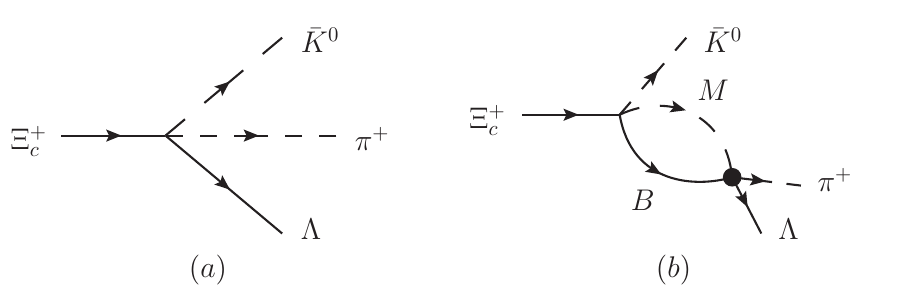}
		\caption{Tree-level diagram (a) and the $S=-1$ final state interaction diagram (b) for the process $\Xi_c^+\to\Lambda\bar{K}^0\pi^+$.}
		\label{fig:hadron1430}
	\end{figure}
 Combining the pseudoscalar mesons matrix and the baryons matrix of Eq.~\eqref{eq:MB}, we can obtain the relevant components of the meson-baryon pair at the hadronic level as
	\begin{align}                 
		&\frac{1}{\sqrt{2}}u\left(\bar{u}u+\bar{d}d+\bar{s}s\right)\left(su-us\right)\\\nonumber
		&=\frac{1}{\sqrt{2}}\sum M_{1i}B_{i2}\nonumber\\
		&=\frac{1}{\sqrt{2}}\pi^0\Sigma^+-\frac{1}{\sqrt{2}}\pi^+\Sigma^0+\frac{1}{\sqrt{6}}\pi^+\Lambda.
		\label{eq:1430MBchannel}
	\end{align}
 Here, we neglect the coupled channels $\eta\Sigma^+$ and $K^+\Xi^0$, since their thresholds are far from the $\Sigma(1/2^-)$ region of interest , and consider the tree-level diagram and the meson-baryon final state interaction, as shown in Fig.~\ref{fig:hadron1430}. On the other hand, as indicated by Eq.~\eqref{eq:1690MBchannel}, the external $W^+$ excitation mechanism can also hadronize into $-\frac{1}{\sqrt{2}}\pi^+\Sigma^0$ and $\frac{1}{\sqrt{6}}\pi^+\Lambda$ meson-baryon pairs, which must likewise be taken into account. Therefor, based on the final states shown in Eq.~(\ref{eq:1430MBchannel}) and Eq.~\eqref{eq:1690MBchannel}, the corresponding decay amplitude for the $\Xi_c^+ \to  \Lambda{\bar{K}}^0\pi^+$ with the transition of the channels to $\pi^+\Lambda$ can be expressed as
	\begin{align}
		\mathcal{T}^{\Sigma(1/2^-)}=&V_p\left[\frac{1}{\sqrt{6}}+\frac{1}{\sqrt{2}}G_{\pi^0\Sigma^+}(M_{\pi^+\Lambda})t_{\pi^0\Sigma^+ \to \pi^+\Lambda}(M_{\pi^+\Lambda})\right. \nonumber \\
            &\left.-\frac{1}{\sqrt{2}}G_{\pi^+\Sigma^0}(M_{\pi^+\Lambda})t_{\pi^+\Sigma^0 \to \pi^+\Lambda}(M_{\pi^+\Lambda})\right. \nonumber \\
		&\left.+\frac{1}{\sqrt{6}}G_{\pi^+\Lambda}(M_{\pi^+\Lambda})t_{\pi^+\Lambda \to \pi^+\Lambda}(M_{\pi^+\Lambda})\right]\nonumber \\
       &+CV_p\left[-\frac{1}{\sqrt{2}}G_{\pi^+\Sigma^0}(M_{\pi^+\Lambda})t_{\pi^+\Sigma^0 \to \pi^+\Lambda}(M_{\pi^+\Lambda})\right. \nonumber \\
   &\left.+\frac{1}{\sqrt{6}}G_{\pi^+\Lambda}(M_{\pi^+\Lambda})t_{\pi^+\Lambda \to \pi^+\Lambda}(M_{\pi^+\Lambda})\right].
		\label{eq:MBchannel}
	\end{align}
With the isospin triplets $(-\pi^+, \pi^0, \pi^-)$ and $(-\Sigma^+, \Sigma^0, \Sigma^-)$, we obtain
		\begin{align}    
	|\pi^0\Sigma^+\rangle	&=-|1,1;0,1\rangle=-\left(-\frac{1}{\sqrt{2}}|\pi\Sigma\rangle^{I=1}+\frac{1}{\sqrt{2}}|\pi\Sigma\rangle^{I=2}\right), \nonumber  \\   
	|\pi^+\Sigma^0\rangle	&=-|1,1;1,0\rangle=-\left(\frac{1}{\sqrt{2}}|\pi\Sigma\rangle^{I=1}+\frac{1}{\sqrt{2}}|\pi\Sigma\rangle^{I=2}\right), \nonumber  \\       
	|\pi^+\Lambda\rangle	&=-|1,0;1,0\rangle=-|\pi\Lambda\rangle^{I=1}.
	\end{align}	
Then we can relate the transition amplitudes of the charge basis with the ones of the isospin basis as follows,
	\begin{gather}
    t_{\pi^0\Sigma^+ \to \pi^+\Lambda} = \langle \pi^0\Sigma^+ | \pi^+\Lambda \rangle = -\frac{1}{\sqrt{2}}t_{\pi\Sigma \to \pi\Lambda}, \\
    t_{\pi^+\Sigma^0 \to \pi^+\Lambda} = \langle \pi^+\Sigma^0 |\pi^+\Lambda \rangle = \frac{1}{\sqrt{2}}t_{\pi\Sigma \to \pi\Lambda}, \\
    t_{\pi^+\Lambda \to \pi^+\Lambda} = \langle \pi^+\Lambda | \pi^+\Lambda \rangle = t_{\pi\Lambda \to \pi\Lambda}.
\end{gather}

    The detailed calculations of the meson-baryon loop function $G$ and the transition amplitude $t$ are the same as in Sec.~\ref{susec:A}, where the subtraction constants are set to $a_{\bar{K} N}=-2.18408$, $a_{\pi\Sigma}=-1.444$, and $a_{\pi\Lambda}=-2.04777$, and the values of the coefficients $C_{ij}$ in the interaction potential are given in Table~\ref{tab:II}, as used in Ref.~\cite{Roca:2013cca}.
	
		\begin{table}
		\renewcommand{\arraystretch}{1.5}  
		\caption {$C_{ij}$ coefficients in the potential  for the coupled channels with $S=-1$ ($C_{ij}$=$C_{ji}$) ~\cite{Roca:2013cca}.}	\label{tab:II}
		\begin{tabular}{lccc}
			\hline\hline  
			\qquad\quad&$\bar{K} N$\qquad\quad&$\pi\Sigma$\qquad\quad&$\pi\Lambda$   \\  \hline 
			$\bar{K} N$\qquad\qquad&0.85\qquad\qquad &-0.93 \qquad\qquad &$-\sqrt{\frac{3}{2}}\times1.056$\\ $\pi\Sigma$\qquad\qquad&\qquad\qquad\qquad\qquad &1.54\qquad\qquad&0    \\
			$\pi\Lambda$\qquad\qquad&\qquad\qquad\qquad\qquad &\qquad\qquad \qquad\qquad &0\\
			
			\hline\hline
		\end{tabular}
	\end{table}

	\subsection{Invariant mass distributions}\label{susec:C}
	
	Combined with the theoretical mechanism given above, one can write down the invariant mass distributions of the double differential width as
	\begin{eqnarray}
		\frac{d^2\Gamma}{dM^2_{\pi^+\Lambda}{dM^2_{\bar{K}^0\Lambda}}}&=&\frac{4M_{\Xi_c^+}M_\Lambda}{{(2\pi)}^3{32M_{\Xi_c^+}}^3}|\mathcal{T}^{\Xi(1/2^-)} + \mathcal{T}^{\Sigma(1/2^-)}|^2. \label {eq:dgammadm12dm23} \nonumber \\
	\end{eqnarray}
 For a given value of  $M_{12}$, the range of $M_{23}$ is defined as,
		\begin{eqnarray}
		M^{\rm max}_{23} &= &\sqrt{\left(E_{2}^\ast+E_{3}^\ast\right)^2 -\left(\sqrt{E_{2}^{\ast2}-m_{2}^2}-\sqrt{E_{3}^{\ast2}-m_{3}^2}\right)^2}, \nonumber \\
		M_{23}^{\rm min} &=&\sqrt{\left(E_{2}^\ast+E_{3}^\ast\right)^2 -\left(\sqrt{E_{2}^{\ast2}-m_{2}^2}+\sqrt{E_{3}^{\ast2}-m_{3}^2}\right)^2},\nonumber \\
	\end{eqnarray}
	where $E_{2}^\ast$ and $E_{3}^\ast$ are the energies of particles 2 and 3 in the $M_{12}$ rest frame, which are written as
		\begin{align}
		&E_{2}^\ast=\frac{M_{12}^2-m_{1}^2+m_{2}^2}{2M_{12}},  \nonumber \\
		&E_{3}^\ast=\frac{M_{\Xi_c^+}^2-M_{12}^2-m_{3}^2}{2M_{12}},
	\end{align}
	where $m_1$, $m_2$, and $m_3$ are the masses of particles 1, 2, and 3, respectively. Finally, by integrating Eq.~\eqref{eq:dgammadm12dm23}, we can easily obtain the invariant mass distributions of the final states $\bar{K}^0\Lambda$ and $\pi^+\Lambda$.

	\section{Results and Discussion} \label{sec:Results}

 Although the branching fraction and the invariant mass distributions of the process $\Xi_c^+ \to \Lambda {\bar{K}}^0 \pi^+$ have not been measured in experiment~\cite{ParticleDataGroup:2024cfk}, its branching fraction of this process without resonant contributions is predicted to be $(4.6\pm 1.2)\%$ by Ref.~\cite{Geng:2018upx}, which implies that this process should be easily measured by Belle II, LHCb, and also the proposed Super Tau-Charm Facility (STCF)~\cite{Lyu:2021tlb}. In this section, we will present our results up to an arbitrary normalization, by taking the global factor $V_p=1$, which does not affect the shape of the invariant mass distribution of the final state.
	
		\begin{figure}[htbp]
		\centering
		\includegraphics[scale=0.58]{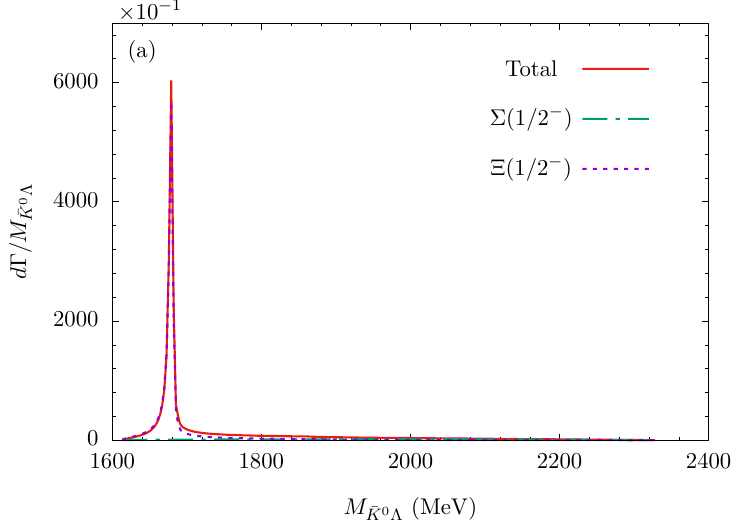}
		\includegraphics[scale=0.58]{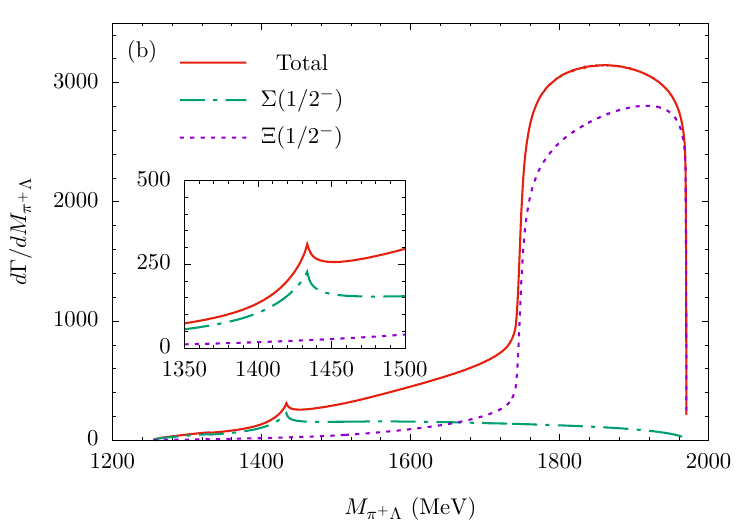}
		\caption{Distributions of $\bar{K}^0\Lambda$ (a) and $\pi^+\Lambda$ (b) invariant masses of the process $\Xi_c^+ \to \Lambda {\bar{K}}^0 \pi^+$.   The red-solid curves show the total contributions, and the green-dot-dashed curves and the purple-dotted curves are the contributions from the  resonances $\Sigma(1/2^-)$ and $\Xi(1/2^-)$, respectively. }	\label{fig:dgdm}
	\end{figure}

       \begin{figure}[htbp]
		\centering
        \includegraphics[scale=0.9]{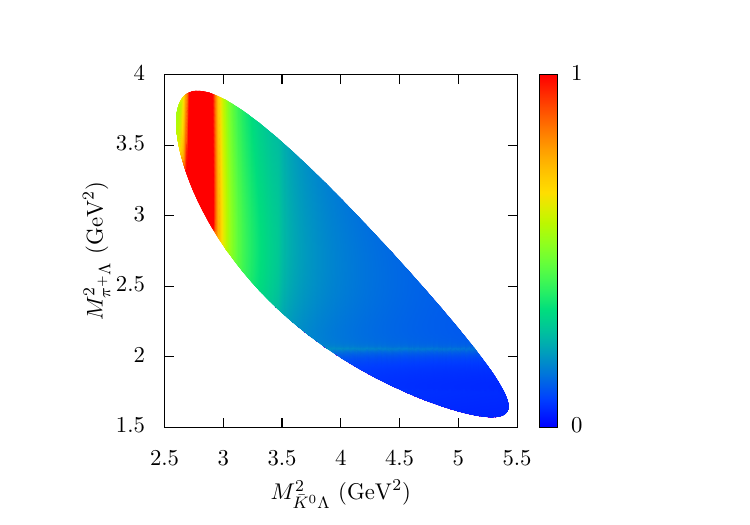}
		\caption{Dalitz plot of $M^2_{\bar{K}^0\Lambda}$ vs. $M^2_{\pi^+\Lambda}$ for the process $\Xi_c^+ \to \Lambda {\bar{K}}^0 \pi^+$}	\label{fig:dalitz}
	\end{figure}

    \begin{figure}[htbp]
	\centering
 
        \includegraphics[scale=0.58]{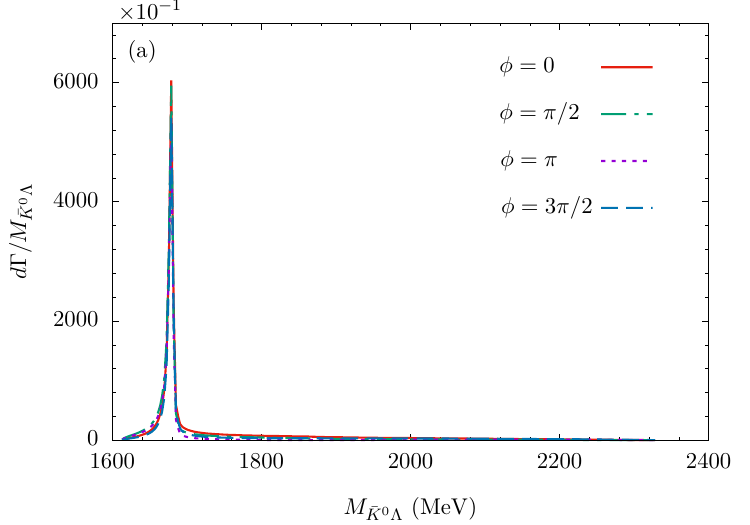}
		\includegraphics[scale=0.58]{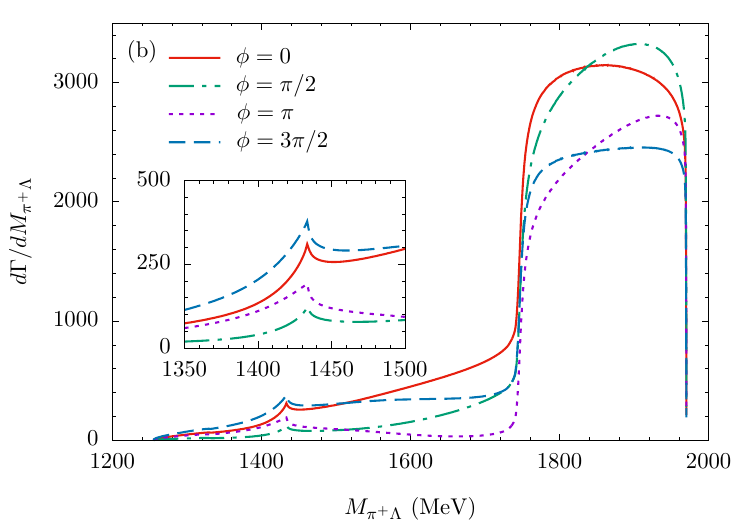}
		\caption{Distributions of $\bar{K}^0\Lambda$ (a) and $\pi^+\Lambda$ (b) invariant masses of the process $\Xi_c^+ \to \Lambda {\bar{K}}^0 \pi^+$ decay with the interference phase $\phi=0, \pi/2, \pi, 3\pi/2$. }	\label{fig:dgdmfie}
	\end{figure}

 \begin{figure}[htbp]
	\centering
 
        \includegraphics[scale=0.58]{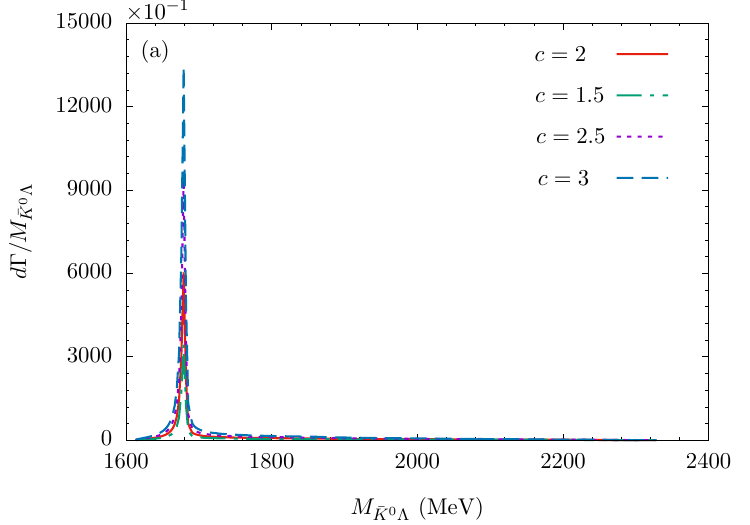}
		\includegraphics[scale=0.58]{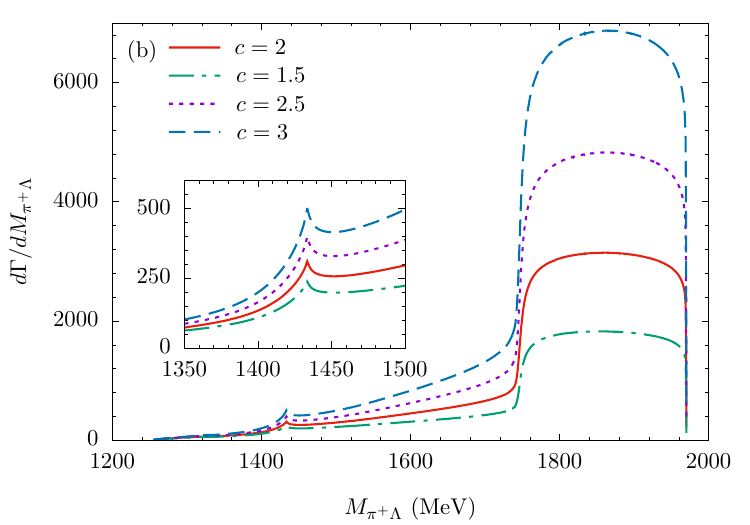}
		\caption{Distributions of $\bar{K}^0\Lambda$ (a) and $\pi^+\Lambda$ (b) invariant masses of the process $\Xi_c^+ \to \Lambda {\bar{K}}^0 \pi^+$ decay with the color factor $C=1.5, 2, 2.5, 3$. }	\label{fig:dgdmcolor}
	\end{figure}

In our model, the color factor $C$ represents the relative weight of the external $W^+$ excitation mechanism with respect to the internal  $W^+$ excitation mechanism in weak interactions. In this work, we take $C=2$ and discuss the dependence of the results on the value of $C$ in the following analysis.

    In Fig.~\ref{fig:dgdm}, we present the $\bar{K}^0\Lambda$ and $\pi^+\Lambda$ invariant mass distributions for the process $\Xi_c^+ \to \Lambda {\bar{K}}^0 \pi^+$, where the red-solid curves show the contribution from the total amplitudes, and the green-dot-dashed and purple-dotted curves represent the contributions from the intermediate $\Sigma(1/2^-)$ and $\Xi(1/2^-)$, respectively. One can find a peak structure in the $\bar{K}^0\Lambda$ invariant mass distribution, which is associated with the resonance $\Xi(1/2^-)$. Additionally, in the $\pi^+\Lambda$ invariant mass distribution, a cusp structure is also observed around 1430~MeV, which should be associated with the predicted $\Sigma(1/2^-)$.  As pointed out in Refs.~\cite{Roca:2013cca,Wang:2015qta}, the predicted $\Sigma(1430)$ couples strongly to the channel $\bar{K}N$, and weakly to the channels $\pi\Sigma$ and $\pi\Lambda$, the strength of the cusp structure in the $\pi^+\Lambda$ invariant mass distribution is not very large because the channel $\bar{K}N$ is not present in the preliminary hadroization process.
    It should be stressed that both $\Xi(1/2^-)$ and $\Sigma(1/2^-)$ are dynamically generated by the $S$-wave pseudoscalar meson-octet baryon interactions, and the corresponding lineshapes of the signals are different with the ones of the Breit-Wigner form.

Next, in Fig.~\ref{fig:dalitz} we show the Dalitz plot for the process $\Xi_c^+ \to \Lambda {\bar{K}}^0 \pi^+$ in the $(M^2_{\bar{K}^0\Lambda}, M^2_{\pi^+\Lambda})$ plane, which corresponds to the doubly differential decay width $d^{2}\Gamma/(dM^2_{\bar{K}^0\Lambda} dM^2_{\pi^+\Lambda})$. In this plot, a significant band corresponding to the $\Xi(1/2^-)$ resonance is clearly visible, and the signal of $\Sigma(1/2^-)$ could also be found. 

Furthermore, the amplitudes of  $\mathcal{T}^{\Xi(1/2^-)}$ and $\mathcal{T}^{\Sigma(1/2^-)}$ could have interference, thus we introduce a relative phase $e^{i\phi}$ between them as follows,
	\begin{align}
		 \mathcal{T}&=\ \mathcal{T}^{\Xi(1/2^-)}e^{i\phi}+\mathcal{T}^{\Sigma(1/2^-)}.
    \end{align}   
Now, we take different values of the phase angle $\phi$ to calculate the invariant mass distributions, as shown in Fig.~\ref{fig:dgdmfie}. One can find that, although the lineshapes of the invariant mass distributions have some changes, both the signals of $\Xi(1/2^-)$ and $\Sigma(1/2^-)$ remain very clear. Thus, the measurements of this process in the future could be used to test the existence of both states, and determine their properties.

In addition, Fig.~\ref{fig:dgdmcolor} illustrates the invariant mass distributions of $\bar{K}^0\Lambda$ and $\pi^+\Lambda$ with different values of color factor $C$, the relative weight of the external excitation with respect to the internal excitation. One can find that the value of $C$ only affects the strength of $\Xi(1/2^-)$ and $\Sigma(1/2^-)$ signals, and their structure are still very clear in the invariant mass distributons.

\begin{figure}[tbhp]
		\centering
		\includegraphics[scale=0.6]{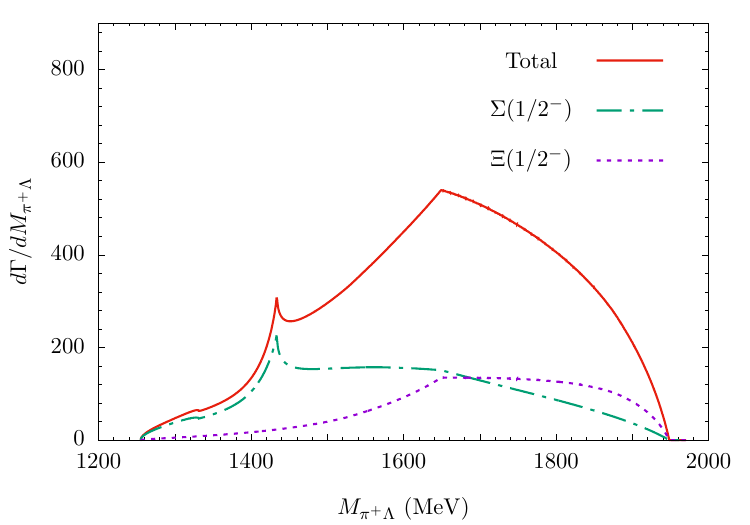}
		\caption{Distribution of  $\pi^+\Lambda$ invariant mass  of the $\Xi_c^+\to\Lambda\bar{K}^0\pi^+$ decay with the cuts of $M_{\bar{K}^0\Lambda}\geq 1750$~MeV.}
		\label{fig:cut off}
	\end{figure}
Since the state $\Xi(1/2^-)$ mainly contributes the energy region of $M_{\pi^+\Lambda}>1.75$~GeV, we take a cut of $M_{\bar{K}^0\Lambda}>
1750$~MeV to exclude the contribution of $\Xi(1/2^-)$, and show the $\pi^+\Lambda$ invariant mass distribution of this process in Fig.~\ref{fig:cut off}. In this figure, a more significant of the cusp structure can be found around 1430~MeV, which should be associated with the predicted $\Sigma(1/2^-)$.


	\section{Summary} \label{sec:Conclusions}
    Since searchES for the low-lying baryons $\Xi(1/2^-)$ and $\Sigma(1/2^-)$ are crucial to deepening our understanding of the light baryons with $J^P=1/2^-$, in this work we have investigated the Cabibbo-favored process $\Xi_c^+ \to \Lambda {\bar{K}}^0 \pi^+$ by considering the $S$-wave pseudoscalar meson and octet baryon interactions within the chiral unitary approach, which can dynamically generate the intermediate states $\Xi(1/2^-)$ and $\Sigma(1/2^-)$. Meanwhile, the contribution from the excited kaons is double Cabibbo-suppressed, and the contribution from the $\Sigma(1385)$ is suppressed due to the Korner-Pati-Woo theory, thus their contributions are expected to be small, and neglected in this work.
    
    	According to our results, one can find a clear peak around 1690~MeV in the $\bar{K}^0\Lambda$ invariant mass distribution and a cusp structure around 1430~MeV in the $\pi^+\Lambda$ invariant mass distribution, which could be associated with the predicted resonances $\Xi(1/2^-)$ and $\Sigma(1/2^-)$. Furthermore, we have shown the Dalitz plot and the dependence of the relative phase angle between the amplitudes of $\Xi(1/2^-)$ and $\Sigma(1/2^-)$, and discussed the influence from the different values of color factor $C$, which implies that the signals of the $\Xi(1/2^-)$ and $\Sigma(1/2^-)$ are alwarys clear in the invariant mass distributions.
    	
Although the branching fraction and the invariant mass distributions of the process $\Xi_c^+ \to \Lambda {\bar{K}}^0 \pi^+$ have not been measurements experimentally, the branching fraction of this process without resonant contributions is predicted to be large, $(4.6\pm 1.2)\%$, by Ref.~\cite{Geng:2018upx}, thus a large sample is expected to be available at Belle II, LHCb, and also the proposed STCF. Meanwhile, the upgrades of BEPCII will increase the collision energy up to 5.6~GeV, which implies that the BESIII Collaboration could perform the measurements on the $\Xi_c$ decays in future~\cite{BESIII:2020nme}.
Therefore we make a call for a precise measurement of this process in these experiments.

\section*{Acknowledgments}
We would like to acknowledge the fruitful discussions with Prof. Eulogio Oset. 
This work is partly supported by  the National Key R\&D Program of China under Grand No. 2024YFE0105200, the Natural Science Foundation of Henan under Grant No. 232300421140 and No. 222300420554, the National Natural Science Foundation of China under Grant No. 12475086, No. 12205075, and No. 12192263. This work is also supported by the Scientific research project of Hunan Provincial Department of Education under Grant No. 24B0106.
	
	
\end{document}